# Dispersive resonance bands within the space charge layer

# of metal- semiconductor junction

S. -J. Tang[1,2], Tay- Rong Chang[1], Chien-Chung Huang[1], Chang-Yeh Lee[2],

Cheng-Maw Cheng[2], Ku-Ding Tsuei[2,1], H. -T. Jeng[3,1], and Chung-Yu Mou[1,4]

*(1) Department of Physics, National Tsing Hua University, Hsinchu 30013 ,Taiwan*
*(2) National Synchrotron Radiation Research Center, Hsinchu 30076, Taiwan*
*(3) Institute of Physics, Academia Sinica, Nankang, Taipei 11529, Taiwan*
*(4) Physics Division, National Center for Theoretical Sciences, P.O.Box 2-131,*
    *Hsinchu, Taiwan*

Based on measurements of angle resolved photoemission, we report that in the

Pb/Ge(111)-$\sqrt{3} \times \sqrt{3}$ R30° structure, in addition to three bands resembling Ge heavy hole (HH), light hole (LH), and split off (SO) bulk band edges, a fourth dispersive band resembling the non split off (NSO) band is found near the surface zone center.

While three Ge bulk-like bands get distorted due to strong coupling between Pb and

Ge, the NSO-like band gets weaker and disappears for larger thickness of Pb, which,

when combined with ab initio calculations, indicates its localized nature within space charge layer. Our results are clearly important for designing electronics involved with metal-semiconductor contacts.

The metal-semiconductor contacts are crucial in every semiconductor device since they provide communication of the device to the outside world. Depending on interface characteristics, the metal-semiconductor contacts may cause different current-voltage behavior [1]. One of the key factors that determine the interface characteristics is the electronic structure at the interface. It is therefore of importance to examine the electronic structure at the interface.

It is known that even a few metal atoms adsorbed on semiconductor surface can induce surface superstructures or reconstructions. The reconstruction of interface may modify the band bending at the interface [2], induce a two-dimensional phase transition [3] or change the properties of overlayer films on top [4]. Another way to change the electronic structure of the interface is to dope the semiconductor. For an n-type semiconductor surface with surface states, if the surface state is acceptor type, space charge effect would cause an upward band bending of the bands near the surface. It thus allows the surface state band to cross the Fermi level so that the bulk donor states would be lifted above the Fermi level. As a result, a positive space charge layer—the depletion layer is built up within a certain depth near the surface. In this case, the density distribution of space charge, $\rho_{sc,}$ is related to the curvature of the band bending via the Poison's equation. For a substantial band bending, the space charge density, $\rho_{sc,}$ can be approximated by a step function with $\rho_{sc}(z) = eN_D^+ \approx eN_D = eN_{SS}/d$, where $N_D$ is the density of bulk donor, $N_{ss}$ is the density of the surface states and $d$ is the depth of the depletion layer. The solution to the Poison's equation with such approximation leads to the magnitude of band bending $V_s$ at the surface [5,6] given by

$$|V_s| = \frac{eN_D d^2}{2\varepsilon\varepsilon_0} \dots\dots\dots\dots\dots\dots\dots\dots\dots\dots\dots\dots(1)$$

Therefore, roughly $d^2 \propto \frac{1}{N_D}$, so the band bending slope or potential gradient would be much higher for the highly doped semiconductor. Fig. 1(a) and (b) shows the comparison of such a behavior for highly doped and lightly doped n- type semiconductors respectively. For the case of the former, the depletion layer is called the inversion layer due to the fact the charge density of holes turns to exceed that of free electrons at the depth near surface. For the highly doped semiconductor, the surface state bandwidth, as shown by the length of the shaded rectangular in Fig. 1(a), could slightly increase with increasing doping density as a result of the fact that donor electrons in the conduction band reduce their energies by occupying the acceptor-like surface states. Consequently, the higher upward band bending of the valence band maximum and the increment of surface state bandwidth would cause overlap between the valence band maximum and the surface state in (E, k) space, hence causing the strong interaction between them.

Experimentally, direct determination of electronic dispersion at the interface has been lacking until in 2005, Takeda *et. al.* first measured quantum well states for the n-type Si quantum well within the In/Si(111)-$\sqrt{7}\times\sqrt{3}$ surface structure by angle resolved photoemission [7]. The in-plane dispersions of the hole subbands in a highly doped semiconductor was first observed. Later, Nathan *et. al.* [8] further discovered fine-structured electronic fringes near the silicon valence band edge in atomically uniform Ag films grown on highly doped n-type Si(111). These fringes were attributed to the quantum slope potential in the inversion layer. Nevertheless, to date, there has been no experimental observation of the signature as a result of large overlap between surface states and the valence band maximum at the surface or interface, as the case shown in Fig. 1(a).

In this paper, based on measurements of angle resolved photoemission, we use highly doped n-type Ge(111) wafer and investigate the interface electronic structure for the Pb/Ge(111)-$\sqrt{3}\times\sqrt{3}$ R30° structure. In addition to three bands resembling Ge bulk band edges, a fourth dispersive band resembling the non split off (NSO) band is found near the surface zone center. The evolution of these four bands with increasing thickness of Pb films at 2M, 4ML and 6ML was also examined. By further comparison with first-principles calculations, we show that the largest weight of the NSO like band is located near the interface of Pb and Ge, confirming the localized nature of NSO like band.

In our study, angle-resolved photoemission measurements were performed by Scienta R3000 energy analyzer with a He I UV light of photon energy 21.2 eV. Some addition data were taken by the 21B1-U9 beamline in the National Synchrotron Radiation Research Center in Taiwan; the results were consistent. The highly doped n-type Ge(111) wafer, $\sim10^{18}$ (1/cm$^3$), was used. The clean procedure for a Ge(111)-c(2x8) surface was described elsewhere [9]. For the formation of Pb/Ge(111)-$\sqrt{3}\times\sqrt{3}$ R30° structure, ~ 3 ML of the Pb film was deposited and the substrate was subsequently annealed from RT to 400 °C. The Pb/Ge(111)-$\sqrt{3}\times\sqrt{3}$ R30° structure thus obtained is β phase as there was no phase transition [10] observed when it was cooled down to -150° C for photoemission measurement. Overlayer uniform Pb film was formed by depositing Pb onto the Pb/Ge(111)-$\sqrt{3}\times\sqrt{3}$ R30° structure kept at -150° C. The calibration of film thickness was done by monitoring the intensity and energy positions of Pb quantum well state (QWS) peak at normal emission as a function of deposition time [11].

To unravel nature of the observed electronic structures, we also perform the first-principles calculations. Our calculations were carried out using the highly accurate full-projected augmented wave method [12] as implemented in the VASP package [13] based on the the local density approximation. The lattice structure of bulk Ge was optimized using a 13x13x7 Monkhorst-Pack k-points mesh over the Brillouin zone with cutoff energy of 500 eV. The Ge(111)-1x1 unreconstructed surface was simulated by a 36-layer Ge slab (thickness of ~ 60 Å) with a vacuum thickness larger than 10 Å, well separating the slabs. The self-consistent calculations were performed over 21 k points mesh over the two-dimensional irreducible Brillouin zone using 27266 plane waves with a cutoff energy of 500 eV under geometry optimization. The long-layer terminated surface was employed to simulate the case under study, in which Pb atoms are believed to mix with the Ge atoms of the top short layer. Based on the optimized structure, the spin-orbital coupling was included in the band structure calculations both for the bulk and surface.

In Fig. 2, we show the gray-scale representations of the photoemission results for the clean Ge(111) -c(2x8) surface (top), and Pb/Ge(111)-$\sqrt{3} \times \sqrt{3}$ R30° surface (bottom), respectively in two symmetry directions . Note that symbols on the top of the Fig. 2 represent the symmetry points for (1x1) surface brillouin zone, while symbols at the bottom represent the symmetry points for $\sqrt{3} \times \sqrt{3}$ R30° surface brillouin zone. It is clear that in the photoemission results either for Ge(111)-c(2x8) or for Pb/Ge(111)-$\sqrt{3} \times \sqrt{3}$ R30°, dispersive energy bands are observed about the symmetry points corresponding to the surface zone boundaries, as indicated by vertical bars. These bands are likely to originate from the surface structures. Even though it is still of interest to investigate in detail those surface electronic bands for both surfaces, and compared them with previously measured and calculated results [10,14-18], the focus of this paper, nevertheless, is on the change of energy band dispersions about the surface zone center $\bar{\Gamma}$ from Ge(111)-c(2x8) to Pb/Ge(111)-$\sqrt{3} \times \sqrt{3}$ R30°. Clearly, Fig. 2 shows a set of four new bands near the surface zone center in both symmetry directions in the Pb/Ge(111)-$\sqrt{3} \times \sqrt{3}$ R30° surface. These four new bands have neither correspondence to the energy bands in Ge(111) surface nor the energy bands in Pb/Ge(111)-$\sqrt{3} \times \sqrt{3}$ R30°, measured or calculated previously [10,14-18]. However, at the first sight, those bands resemble the bulk hole band edges of Ge. Since the Ge(111) used in this experiment is highly doped n-type, the four new bands observed might be related to the doping effect from Ge(111). In Fig. 3(a) and (b), we superimposed the calculated Ge bulk band edges dispersing in the same symmetry direction ($\Gamma$K for bulk, $\bar{\Gamma}\bar{K}$ for 1x1, and $\bar{\Gamma}\bar{M}\bar{K}$ for $\sqrt{3} \times \sqrt{3}$ R30°) onto the 2D image data. As is evident, the calculated bulk band edges, HH, LH, SO (red dashed curves), and NSO (blue dot curves), match the measured four bands very well within the large energy range from Fermi level down to 1.0 eV. The photon energy dependent data, which is not presented in this paper, shows the energies of the four bands don't shift with photon energies, indicating the surface related nature. Hence, the scenario represented by Fig. 1(a) seems to apply, in which surface state become a surface resonance (SR), crossing the inversion layer and strongly mixing with bulk band edges. Similar results were observed by Tang *et. al*. [19] that the surface state of Ag thin film at the thickness smaller than its decay length exhibits the energy dispersions resembling HH, LH and SO band edges as a result of the strong interaction with Ge bulk edges.

However, two questions still remain. First, why aren't those four bands observed from a clean highly doped n-type Ge(111) surface? Second, why does the NSO like band exist, which doesn't have the counterpart in the bulk? The surface state bands of clean Ge(111)-c(2x8) about the zone center, as recently reexamined by Razado-Colambo *et. al.* [16], mostly originated from the dangling bonds of rest atoms and adatoms, or back bonds few layers below them near the surface [14-16], which are intimately related to the c(2x8) surface reconstruction. The higher intensities of those localized surface state bands would cover the relatively lower intensities of the four bands [20]. In addition, adsorption of Pb on Ge(111) change the surface reconstruction from c(2x8) to $\sqrt{3} \times \sqrt{3}$ R30°, hence diminishing these Ge(111)-c(2x8) localized surface state bands and facilitating the appearance of the four bands. Fig. 3(b) shows the energy band dispersions for 2ML Pb on Ge(111) and surprisingly, the four bands remain and three bands, corresponding to HH, LH and SO, get even sharper and more intense. The independence of these three bands on the lattice structures above the Ge(111) surface confirms that their wave functions have most of amplitudes below the surface area, going through the inversion layer, as depicted by the bottom of Fig. 1(a). This interesting finding from 2 ML yet points to another role Pb plays, which is to increase the SR density or bandwidth so as to have stronger interaction with Ge bulk band edges even though the SR were originally contributed by the bulk donors in the inversion layer. When Pb film get thicker, the coupling of Ge bulk band edges with SR should transit to the coupling with the quantum well resonances (QWR) of Pb films, and 2ML is at the intermediate position for this transition. As seen from Fig. 3 (c),(d),(g),(h), the curvatures of three bands corresponding to HH, LH, and SO band edges get expanded but the fourth band corresponding to NSO yet get weaker and tend to disappear with increasing thickness from 4ML to 6ML [21]. The density of states for Pb electrons are not dominated by Ge bulk edges any more; instead, the three new bands can be simulated through the Anderson's interaction between discrete QWR and continuous bulk edges [9]. The net effect is like that the Ge band edges split the QWR bands into several segments. For comparison, the expected subbands of QWS for free standing Pb films of 2ML, 4ML and 6ML, based on a tight-binding calculation of the bulk band structure of Pb [22] are superimposed on to the data, as shown by the green curves in Fig. 3(f),(g),(h). The crossover of the QWR band and Ge band edges presents kink or break like structures, which are more evident for 6ML.

Due to the observed fact that the intensity of the NSO like band get weaker and tend to disappear with increasing thickness, the origin of the fourth band should not be due to the coupling between the thin film and substrate, and hence not be related to the valence band bending. The mean free path of the electrons corresponding to the photon energy used (21.2 eV) is about 10 Å, which is about 4 ML of Pb film, wherein the intensity of the fourth band was observed to start decaying. Therefore, it is reasonable to consider it as some unique 2D band, which is inherently localized within the space charge layer. To simulate the hole subband dispersion of the space charge layer, we perform first-principles calculation based on a 36 layer Ge(111) slab [23]. The spectrum is shown in Fig. 4. It is clearly seen that in Fig. 4(a), there is a trace of a SR type band isotropically across the discrete hole subbands in the directions from the surface zone center $\bar{\Gamma}$ to the two surface zone boundaries at $\bar{K}$ and $\bar{M}$, respectively. The trace of the band is extracted as indicated by the blue circles inside the rectangular frame of the Fig. 4(b), and superimposed it onto our 2D photoemission image data for 2ML of Pb film in both symmetry directions within the same range as the rectangular frame. As seen in Fig. 4(c), this calculated SR band matches the fourth band very well, located inside the gap induced by the spin orbital splitting (SOS) of the three hole band edges. According to our calculation, this SR band has the mixture of $p_x$ and $p_y$ symmetry. Fig. 4(d) shows the z profile of the charge density of this SR at $\bar{\Gamma}$. As seen, it has negligible electron density at and above the first layer; instead, 47% and 39% of the electron density are distributed over the second and third layer, and over the fourth and fifth layer, respectively. This result explains why the fourth band, whereas a Ge inherent 2D band, is still observed after Pb deposition because of its delocalization from the surface. Surface states existing within the SOS induced gap have been previously observed on the metal surfaces [24-26]. The existence of such kind of surface state is very likely for Ge surface since Ge has fairly large SOS induced gap of ~ 0.3 eV at the zone center. Nevertheless, this gap is a pseudogap in light of the fact that the hole subbands still exist inside the gap and therefore, the fourth band observed is actually a SR.

In conclusion, we have studied electronic structures of Pb thin films on highly doped n-type Ge(111) surface. Due to the largely reduced depth of the inversion layer, there is a strong electron–electron coupling between the Pb thin film and the Ge hole band edges within the layer. With increasing thickness of Pb film, this coupling changes from the type of SR- bulk edge coupling to the type of QWR- bulk edge coupling. The NSO like band observed from the very thin film is another type of SR band derived from the SOS induced gap. The agreeable match between the measurement and first-principles calculation without assuming the shape of the space charge layer indicates that this NSO band is an intrinsic SR state localized within the space charge layer of Ge(111) surface, regardless of the doping concentration of Ge. However, more delicate and detailed calculation work is indeed needed in the future in order to fully understand the origin of the NSO band. We believe this newly found SR band is very important to the property of semiconductor interface or surface such as the electric conductivity, which has been studied experimentally and theoretically [27,28] mainly from the Si surface that, however, has very negligible SOS induced gap. In addition to that, the spin transport induced by the Rashba splitting at the interface of the semiconductor has gained massive interest recently [29]. Therefore the influence of this SR band on the properties of Ge(111) surface and interface, and its potential application is yet to uncover.

**Acknowledgements:**

The research is supported by the National Science Council of Taiwan (grants NSC98–2112–M-007–017–MY3 and NSC 95-2112-M-007-062-MY3) and by the National Synchrotron Radiation Research Center. The author would like to thank B. -S. Fang for providing the vacuum chamber and instruments in setting up the new angle-resolved photoemission system.

(Plenum Press, New York, 1986).

**Figure Captions:**

1.  The energy positions of valence band maximum ($E_v$), conduction band minimum ($E_c$), intrinsic energy ($E_i$), and Fermi level ($E_f$) as a function of the distance $z$ from the surface of the (a) highly doped and (b) lightly doped n type semiconductor. The bandwidth of occupied (unoccupied) acceptor type surface state is represented by a shaded (blank) rectangular. The bottom panel shows the wave function of the acceptor type surface state corresponding to (a) and (b).

2.  Angle-resolved photoemission data presented as grey-scale images as a function of energy and $k_{||}$ for the clean Ge(111)-c(2x8) surface (top), and Pb/Ge(111)- $\sqrt{3} \times \sqrt{3}$ R30° surface (bottom), respectively in two symmetry directions .

3.  Angle-resolved photoemission data presented as grey-scale images as a function of energy and $k_{||}$ for (a) Pb/Ge(111)- $\sqrt{3} \times \sqrt{3}$ R30°, and (b) 2 ML (c) 4 ML and (d) 6 ML of Pb on Ge(111) in the symmetry direction $\Gamma K$ for bulk, $\overline{\Gamma}\overline{M}\overline{K}$ for $\sqrt{3} \times \sqrt{3}$ R30°, and $\overline{\Gamma}\overline{K}$ for 1x1. The calculated bulk HH, LH, SO (red dashed curves) and NSO (blue dot curves) bands of Ge, and the calculated QWS subbands (green solid curves) of Pb films are superimposed onto the data in (e), (f), (g) and (h), correspondingly.

4.  (a) The hole subband dispersions of Ge(111) space charge layer in the symmetry

directions $\overline{\Gamma}\overline{K}$ and $\overline{\Gamma}\overline{M}$, based on a 36 layer slab model. (b) The trace of SR band, as indicated by the blue circles, extracted from the hole subband dispersions in both symmetry directions. (c) The extracted SR band superimposed on the angle-resolved photoemission data for 2 ML Pb on Ge(111) in both symmetry directions. The red dashed curves are the corresponding calculated HH, LH and SO hole band edges. (d) The charge density z profile for the SR state at $\overline{\Gamma}$. The layer numbers are indicated on the top.

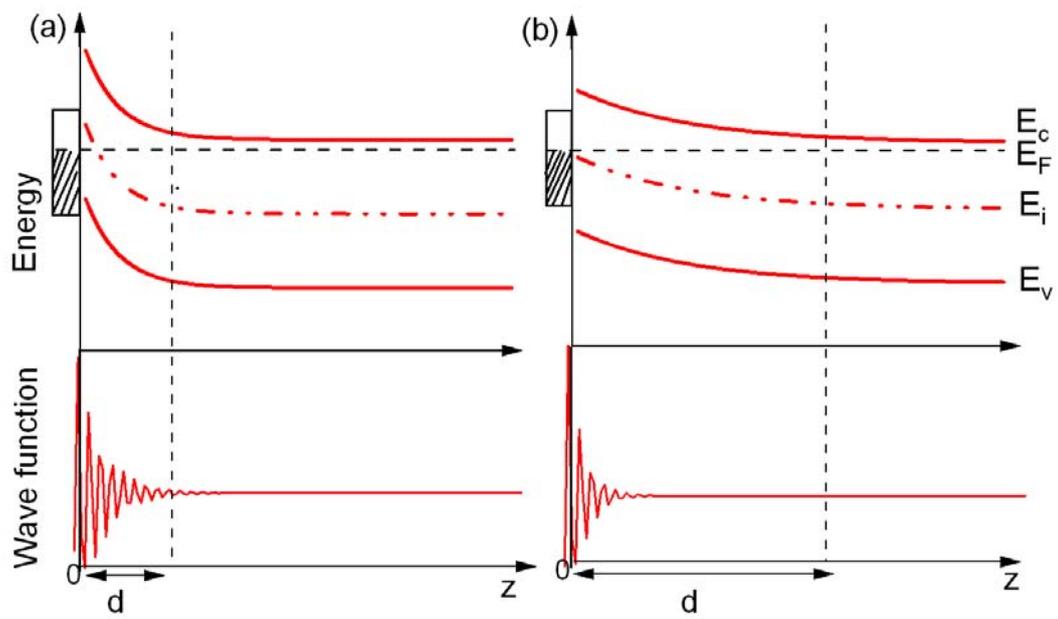

Figure 1

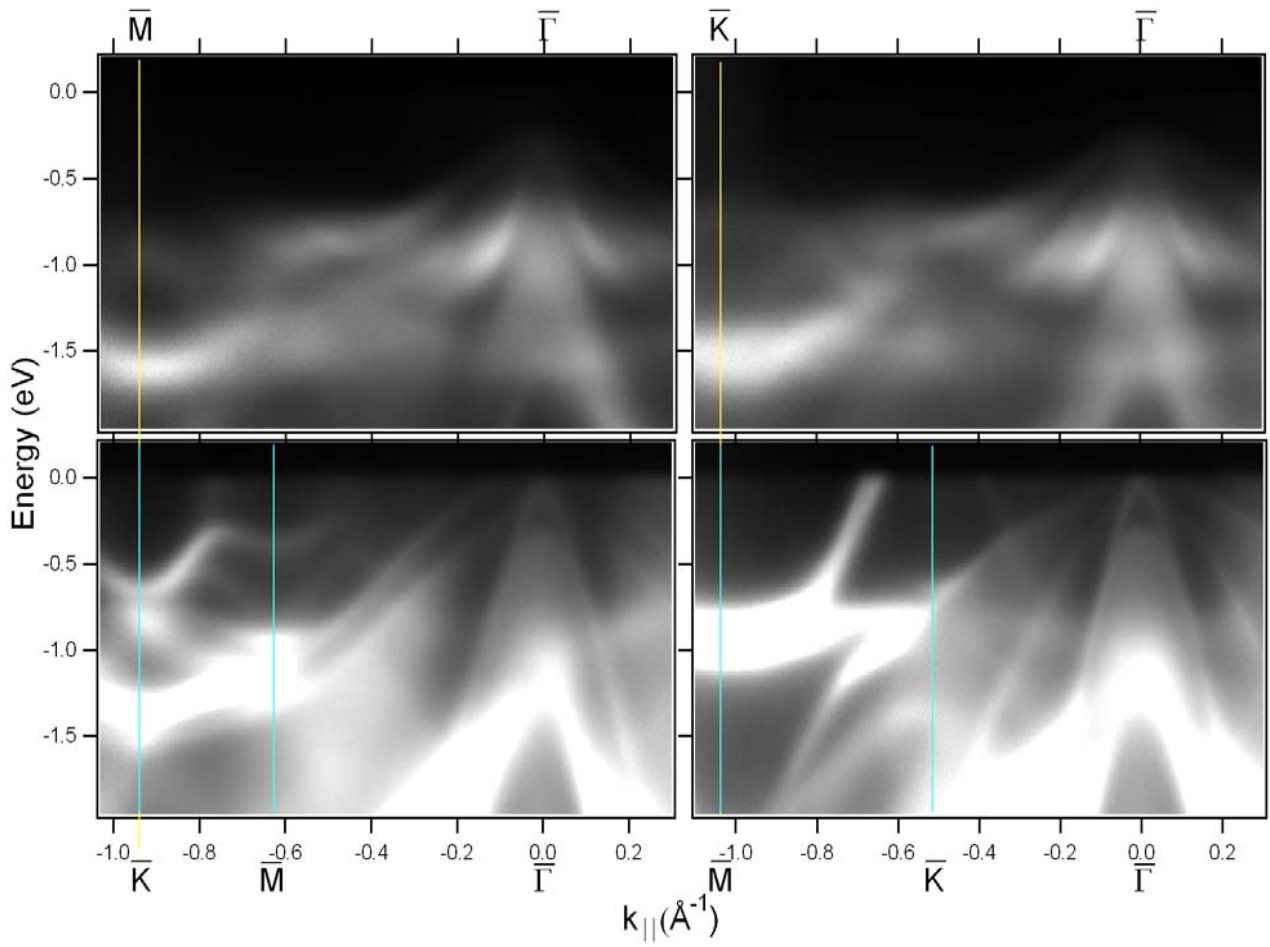

Figure 2

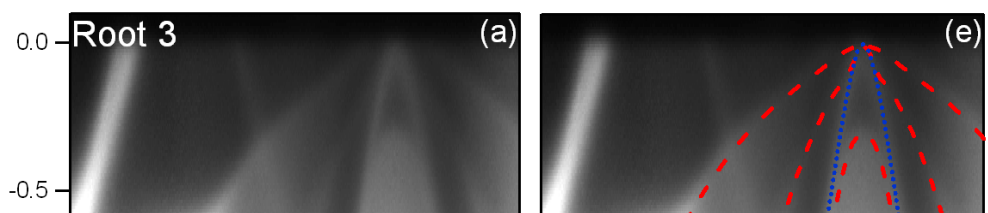

Figure 3

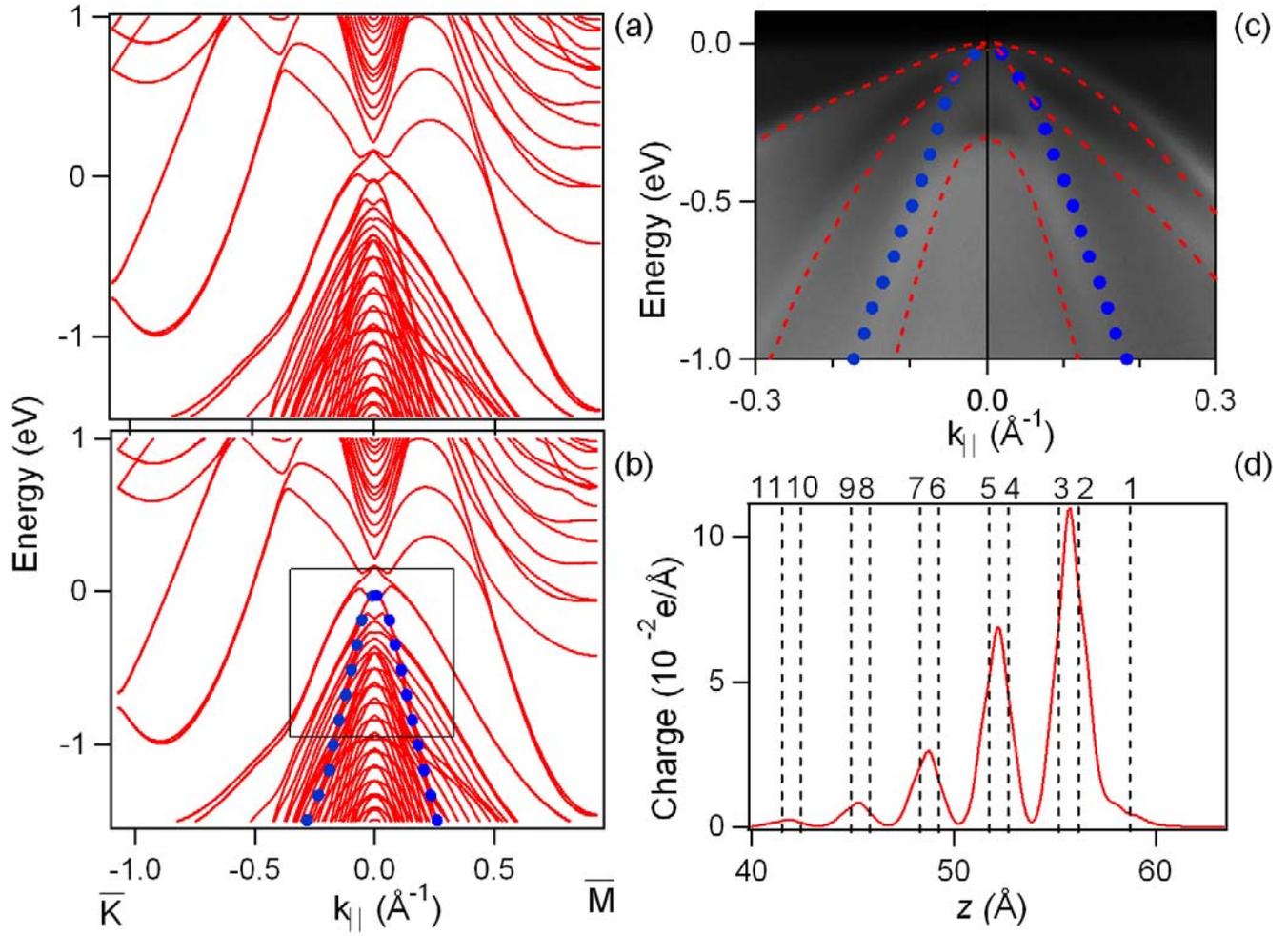

Figure 4